\documentclass[prl,amssymb,twocolumn,tightenlines,showpacs]{revtex4-1}

\usepackage{graphicx}
\usepackage{amsmath}
\usepackage{verbatim}
\usepackage{subfigure}
\usepackage{epstopdf}

\begin{document}

\title{Experimental super-resolved phase measurements with shot-noise sensitivity}

\author{L. Cohen}
\affiliation{Racah Institute of Physics, Hebrew University of
Jerusalem, Jerusalem 91904, Israel}
\author{D. Istrati}
\affiliation{Racah Institute of Physics, Hebrew University of
Jerusalem, Jerusalem 91904, Israel}
\author{L. Dovrat}
\affiliation{Racah Institute of Physics, Hebrew University of
Jerusalem, Jerusalem 91904, Israel}
\author{H. S. Eisenberg}
\affiliation{Racah Institute of Physics, Hebrew University of
Jerusalem, Jerusalem 91904, Israel}


\begin{abstract}
The ultimate sensitivity of optical measurements is a key element
of many recent works. Classically, it is mainly limited by the
shot noise limit. However, a measurement setup that incorporates
quantum mechanical principles can surpass the shot noise limit and
reach the Heisenberg limit. Nevertheless, many of those
experiments fail to break even the classical shot-noise limit.
Following a recent proposal, we present here the results of
optical phase measurements with a photon-number resolving detector
using coherent states of up to 4200 photons on average. An
additional scheme that can be implemented using standard
single-photon detectors is also presented, and the results of the
two schemes are compared. These measurements present deterministic
single-shot sub-wavelength super-resolution up to 288 better than
the optical wavelength. The results follow the classically limited
sensitivity, up to 86 times better than the wavelength.

\end{abstract}

\maketitle

One of the most common ways to sense and measure the physical
world is by using electromagnetic radiation, and in particular
light. Usually, the sensitivity of such measurements is governed
by the quality of the measuring device, but ultimately, this
sensitivity is bounded by fundamental physical limits on the
measurement uncertainty \cite{Meyer88}. The resolution of a
measurement, defined as the width of its smallest detail, follows
in most cases the scale of the light's wavelength $(\lambda)$.
Nevertheless, the sensitivity scale is not the resolution.
Instead, it is determined by the measurement signal-to-noise ratio
and can exceed the resolution by many orders of magnitude. For
classical measurement devices, sensitivity is limited by the shot
noise limit (SNL) -- the discrete division of light energy into
photons and their Poissonian statistics.

Interestingly, both resolution and sensitivity can be further
enhanced. Super-resolved phase measurements, where the smallest
detail is smaller than the Rayleigh limit $(\lambda/2)$, have been
demonstrated with many approaches, using either classical or
quantum light sources. The shot noise limit is surpassed by
super-sensitive methods that incorporate nonclassical states of
light and measurements. As states with non-Poissonian statistics
can be used, sensitivity is only limited by the Heisenberg
uncertainty principle \cite{Giovannetti11}. Most notable is the
use of squeezed states for this purpose \cite{Aasi13}. In the last
decade there has also been a wide interest in the so-called NOON
states \cite{Dowling08}, as they exhibit both super-resolution and
super-sensitivity. These states are superpositions of $N$ photons
in one mode and none in the other, and the opposite arrangement of
no photons in the first and $N$ in the second.

By using NOON states, super resolution has been demonstrated for
states with $N=2$ \cite{Kuzmich98}, $N=3$ \cite{Mitchell04}, $N=4$
\cite{Nagata07}, and $N=5$ \cite{Afek10}. Nevertheless,
sensitivity has only been improved up to 1.59 times better than
the SNL, in a probabilistic manner where the signal is
post-selected from a larger set of outcomes \cite{Nagata07}. Using
hyper-entanglement between ten degrees of freedom of five photons,
10 times resolution enhancement has been demonstrated, as well as
$13\%$ sensitivity enhancement for four photons with eight degrees
of freedom \cite{Gao10a}. Regrettably, despite their advantages,
NOON states are hard to generate, vulnerable to noise, and require
special detection setups.

Unlike NOON states, optical coherent states are generated by
common lasers. They are reduced by absorption to weaker coherent
states, thus keeping their qualitative behavior. Enhanced
resolution can be achieved by using the photons simultaneously in
a few parallel measurements \cite{Kothe}, or consecutively by
recycling them through a series of measurements
\cite{Higgins,Migdall}. The Fabry-P\'{e}rot interferometer is a
standard example for the latter. Alternatively, the measuring
apparatus can measure properties other than just the average
power. Such enhancement has been demonstrated by post-selecting
the projection of the output of an interferometer onto a Fock
state of 7 photons \cite{Khoury}, and onto a NOON state of $N=6$
\cite{Resch07}, and deterministically by homodyne detection
\cite{Distante13}.

In this Letter we present the experimental results of a recently
proposed scheme by Gao \textit{et al.}, that enables the
demonstration of super-resolution, while staying at the SNL
\cite{Gao10b}. It combines standard interferometry of coherent
states with a special nonclassical measurement that requires
photon-number resolving detectors. Additionally, we show how a
similar scheme, that uses only single-photon detectors that cannot
resolve the number of photons, achieves super-resolution which is
only marginally worse than the original proposal, while keeping
the same shot noise limited sensitivity.

The special measurement we applied in this work is the
photon-number parity. The parity operator is defined as
$\hat{\pi}=(-1)^{\hat{N}}$, where $\hat{N}$ is the photon number
operator. Its expectation value is
$\langle\hat{\pi}\rangle=\sum_{n=0}^\infty (-1)^nP_n$, where $P_n$
is the probability for $n$ photons in the detected state. The
parity expectation value reflects the oddness or evenness of the
photon number distribution. For example, the parity of a Fock
state is 1 or -1 for an even or odd $N$, respectively. Parity
measurements were first used with trapped ions \cite{Bollinger},
and later in the context of optical interferometry
\cite{Gerry00,Gerry01}. Apparently, parity measurements are
advantageous for many light sources \cite{Chiruvelli,Gerry10} and
can even break the Heisenberg limit if the light source is a two
mode squeezed vacuum state \cite{Anisimov}. It can be detected
indirectly using homodyne techniques \cite{Plick,Distante13}, or
directly by observing the photon-number distribution with a
photon-number resolving detector, as demonstrated here.

Consider a Mach-Zehnder interferometer whose input is a coherent
state from one port and the vacuum from the other. The relative
phase $\phi$ between its arms is the variable we would like to
measure, while we observe the state of one of the two output
ports. The detected output state is a phase sensitive coherent
state $|\alpha(\phi)\rangle$. The parity expectation value of this
state is \cite{Gao10b}
\begin{equation}\label{PEV}
\langle\alpha|\hat{\pi}|\alpha\rangle=e^{-2|\alpha|^2}=e^{-2\bar{n}\cos^2(\frac{\phi}{2})}\,_{\overrightarrow{\phi\rightarrow\pi}}e^{-\frac{\bar{n}(\phi-\pi)^2}{2}}\,,
\end{equation}
where $\bar{n}=|\alpha(0)|^2$ is the maximal detected average
photon number, which already includes the effects of photon loss
in the interferometer and in the detector. This expectation value
is approaching zero as the average number of output photons and
its distribution width are increased, because wider distributions
include odd and even terms more evenly. On the other hand, when
the average photon number is approaching zero, the parity
approaches one, as the dominant term becomes $P_0$, an even term.
When the resolution is defined as half width at $1/e$ of the
maximum, the parity measurement resolution is
$\sqrt{{2}/{\bar{n}}}$. The parity measurement sensitivity
saturates the SNL \cite{Gao10b,Anisimov}. It is derived using
quantum estimation theory to be
\begin{equation}\label{precisionPi}
\Delta\phi=\frac{\Delta\hat{\pi}}{|\frac{d\langle\hat{\pi}\rangle}{d\phi}|}\,_{\overrightarrow{\phi\rightarrow\pi}}\frac{1}{\sqrt{\bar{n}}}\left(1+(2\bar{n}+1)\frac{(\phi-\pi)^2}{8}\right) \,,
\end{equation}
where the parity uncertainty is
$\Delta\hat{\pi}=\sqrt{1-\langle\hat{\pi}\rangle^2}$
\cite{Khoury}.

The major contribution that the $P_0$ component has on parity at
its peak suggests that this value is responsible for most of the
improvement in resolution and sensitivity. We explore this option
as $P_0$ can also be obtained by a standard single-photon detector
that cannot discriminate between different photon numbers. Its two
outcomes, no photon and any number of photons, corresponds to the
measurement of $P_0$ and $P_r=\sum_{n=1}^\infty P_n$,
respectively. One may assume that the difference between these two
values is a better approximation of parity, as the deviation is
only at the third term $P_2$. Nevertheless, as $P_0-P_r=2P_0-1$,
considering only $P_0$ is equivalent, and much simpler to
evaluate.

The observable for no detection is the zero photon projector
$\hat{Z}=|0\rangle\langle0|$, and its expectation value $P_0$ is
\begin{equation}\label{P0}
\langle\alpha|\hat{Z}|\alpha\rangle=e^{-|\alpha|^2}=e^{-\bar{n}\cos^2(\frac{\phi}{2})}\,_{\overrightarrow{\phi\rightarrow\pi}}e^{-\frac{\bar{n}(\phi-\pi)^2}{4}}\,.
\end{equation}
This is a very similar result to Eq. \ref{PEV}, with a
super-resolved peak which is only wider by $\sqrt{2}$. In order to
find the sensitivity of this measurement, we follow the same
procedure as in Eq. \ref{precisionPi} and obtain
\begin{equation}\label{precisionP0}
\Delta\phi=\frac{\Delta\hat{Z}}{|\frac{d\langle\hat{Z}\rangle}{d\phi}|}\,_{\overrightarrow{\phi\rightarrow\pi}}\frac{1}{\sqrt{\bar{n}}}\left(1+(\frac{\bar{n}}{2}+1)\frac{(\phi-\pi)^2}{8}\right)
\,,
\end{equation}
where we used $\Delta\hat{Z}=\sqrt{P_0(1-P_0 )}$. Surprisingly,
the $P_0$ measurement is also shot-noise limited.

In order to realize the measurements discussed above, we use a
polarization version of the Mach-Zehnder interferometer, where the
two spatial modes are replaced by the horizontal and vertical
polarizations. In order to achieve better visibility, the
polarization modes are not coupled by $\lambda/2$ wave-plates.
Instead, the input and output photons pass through Glan-Thompson
polarizers oriented at $45^\circ$. The relative phase between the
polarization modes is introduced by tilting a calibrated 2\,mm
thick birefringent calcite crystal. An identical crystal oriented
at $90^\circ$ is used to compensate for temporal walk-off. The
output is directed to a silicon photo-multiplier (SiPM) detector,
which is an array of many single-photon detecting elements, and
thus has photon-number sensitivity. The input state is generated
by a Ti:Sapphire laser with 780\,nm wavelength, that produces
150\,fs long pulses. The pulses are picked to reduce their rate to
250\,kHz. The average number of photons per pulse is controlled by
calibrated neutral density filters.

\begin{figure}[b]
\includegraphics[angle=0,width=86mm]{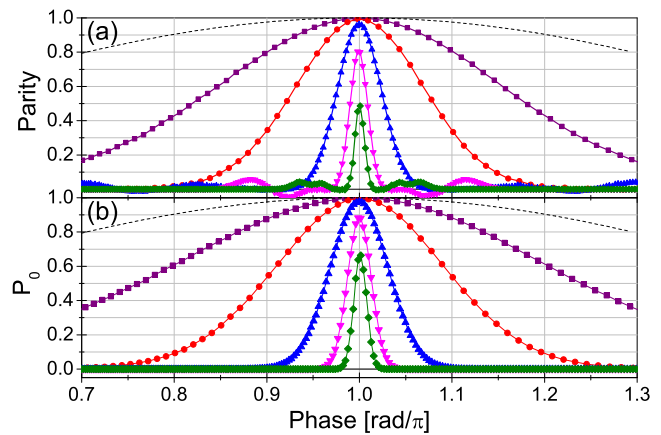}
\caption{\label{Fig1}(color online) Parity (a) and $P_0$ (b)
dependence on the phase difference between the two polarization
modes. The presented measurements are for average photon numbers
of $4.6\pm{0.2}$ (purple squares), $25\pm{1}$ (red circles),
$200\pm{8}$ (blue triangles), $1190\pm{50}$ (pink inverted
triangles),and $4150\pm{150}$ (green rhombuses). The dashed black
lines represents the classic interference curves, for comparison
reasons. Errors are not shown, as they are smaller than the
symbols.}
\end{figure}

There are a few effects that distort the original photon
statistics as registered by SiPM detectors
\cite{Dovrat,DovratSim}. The imperfect detection efficiency
reduces the average detected photon number, but does not affect
its Poissonian statistics. Dark counts are false detections as a
result of thermal excitations. Cross-talk can trigger neighboring
elements of photon triggered elements, and thus falsely increase
the number of detected photons. Finally, there is a chance for
more than one photon to hit the same detecting element as a result
of their finite number (100 in our case). As around the
significant region of our measurement $|\alpha|\ll1$, only dark
counts affect it, and their effect is important only for
relatively small $\bar{n}$.

Figure \ref{Fig1} presents the super-resolved signals obtained for
parity (Fig. \ref{Fig1}a) and $P_0$ (Fig. \ref{Fig1}b). The data
sets are for increasing average photon numbers. A clear narrowing
of the signals is observed, where the narrowest peak was measured
for parity with about 4000 photons. This peak's width is
$\lambda/(288\pm3)$, corresponding to resolution which is 144
times better than what is regularly achieved (presented for
comparison by a black dashed line). The corresponding widths of
the curves for parity and $P_0$ differ by $\sqrt{2}$, as expected
from theory. Another difference between the two results is the
weak oscillations that appear only for parity. They are a result
of the truncation of the parity measurement at 26 photons, which
becomes more significant as more photons are involved.

\begin{figure}
\includegraphics[angle=0,width=86mm]{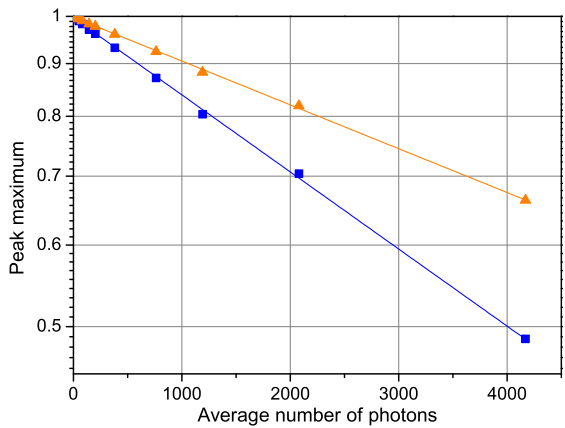}
\caption{\label{Fig4}(color online) The effects of finite
visibility and dark counts on the signal peak heights for parity
(blue squares) and $P_0$ (orange triangles). The peak heights are
decreased when states with more photons are used, according to Eq.
\ref{Heights}.}
\end{figure}

The degradation of the peak heights as the average photon number
is increased, is a result of the imperfect visibility $V$ of the
interferometer, and the dark counts, whose average number is
$n_d$. The visibility is defined as
$V=(\bar{n}-n_b)/(\bar{n}+n_b)$, where $n_b$ is the average number
of background counts at $\phi=\pi$. Thus, it can be shown that the
peak heights $h$ follow
\begin{equation}\label{Heights}
h(\bar{n})=\exp{\left[-\beta\left(n_d+\frac{(1-V)\bar{n}}{2}\right)\right]}\,,
\end{equation}
where $\beta=2$ for parity and 1 for $P_0$. Figure \ref{Fig4}
presents the peak height of all measurements with their fits to
Eq. \ref{Heights}. The data fits theory very well, where the
parity curve slope is about twice as large as that of $P_0$. From
the fit parameters of both curves we extract the dark count rate
to be $400\pm100$\,Hz and the visibility to be
$0.99981\pm10^{-5}$, in agreement with independent direct
measurements.

\begin{figure}
\includegraphics[angle=0,width=86mm]{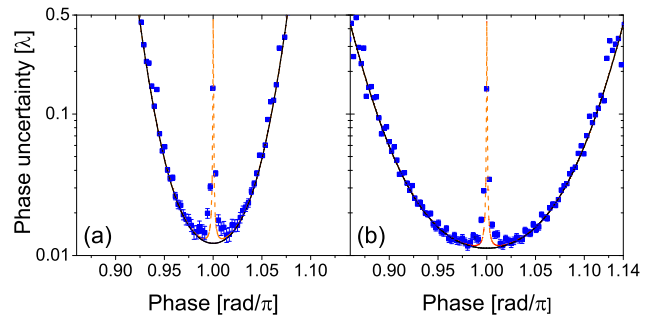}
\caption{\label{Fig2}(color online) Experimental phase
uncertainties for 200 photons on average (blue squares), as a
function of the phase for (a) parity and (b) $P_0$. The
theoretical fits to Eqs. \ref{precisionPi} and \ref{precisionP0}
without approximation are presented with solid black lines. Fits
to equations that include the effect of imperfect visibility are
presented by orange dashed lines. Errors are shown when larger
than their symbols. They represent the standard deviation in the
phase uncertainty estimation process.}
\end{figure}

The experimental single-shot phase uncertainty for parity and
$P_0$ were calculated from their data using Eqs. \ref{precisionPi}
and \ref{precisionP0}, respectively. Representative results for
200 photons on average are presented in Fig. \ref{Fig2}. The
uncertainty for $P_0$ is at minimum for a larger range as its peak
is also wider. The imperfect visibility is also responsible for
the large deviation near $\phi=\pi$ \cite{Kuzmich98}. As the
uncertainty in the estimated quantity is a ratio between the
uncertainty in the measured quantity and its slope, it can only be
finite at this point as long as both contributions approach zero
together. Imperfect visibility results in non-zero observed value
and non-zero uncertainty in it. Its slope on the other hand is
always zero at this point, where this value peaks, resulting in
diverging uncertainty in the estimated value. Fits to modified
equations which include imperfect visibility due to background
counts are also presented, with a good agreement with the results.

\begin{figure}
\includegraphics[angle=0,width=86mm]{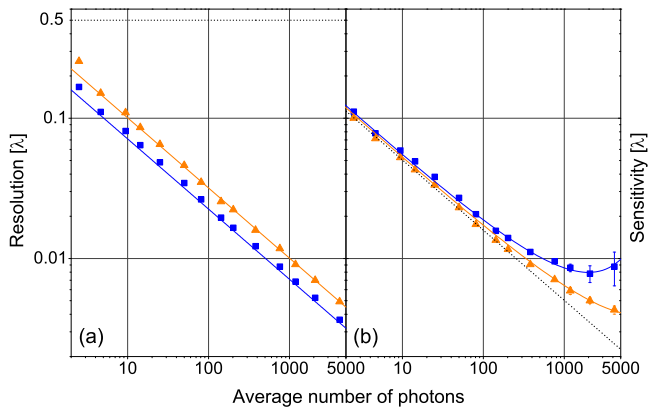}
\caption{\label{Fig3}(color online) (a) The resolution and (b) the
smallest uncertainty for parity (blue squares) and $P_0$ (orange
triangles) measurements. Solid lines are the theoretical
predictions. For the resolution there are no free parameters and
the predictions for sensitivity use the visibility values from the
fits of Fig. \ref{Fig4}. The parity resolution reaches
$\frac{\lambda}{288}$, 144 times better than the Rayleigh
$\lambda/2$ limit (dashed line in (a)). The $P_0$ sensitivity
follows the SNL (dashed line in (b)) up to 200 photons. Parity has
better resolution but larger deviation from the SNL. Errors were
calculated as before and shown when larger than their symbol.}
\end{figure}

A summary of all the results for the range of 2.5 to 4200 photons
on average is presented in Fig. \ref{Fig3}. The resolutions of
parity and $P_0$ are separated by $\sqrt{2}$ as expected. They are
not affected by the imperfect visibility. The phase uncertainty
for $P_0$ is slightly better for small photon numbers, where it
follows meticulously the theoretical SNL line. Actually, within
experimental errors, the sensitivity of the $P_0$ measurement is
shot-noise limited up to the measurement with 200 photons, where
the sensitivity is $\lambda/(86\pm2)$ and the resolution is
$\lambda/(45\pm1)$. Although when more photons are used this limit
is not followed anymore, phase sensitivity is further improved,
until when using 4200 photons it is $\lambda/(230\pm15)$ and the
resolution is $\lambda/(202\pm2)$. The better sensitivity of $P_0$
is explained by the dependence on a single measured value,
compared to parity that depends on many.

We should emphasize that the values we present here assume photon
detection with perfect quantum efficiency. The detector we have
used is far from having a perfect efficiency. Nevertheless, this
is a frequently used practice, that demonstrates the limit of the
method. Moreover, there are other kinds of photon-number resolving
detectors with quantum efficiencies that approach unity
\cite{Waks03, Lita08}. Regrettably, they are not available to us.

In conclusion, using a photon-number resolving detector, we have
measured the parity and the probability for no-detection of a
coherent state that has travelled through a Mach-Zehnder
interferometer. Every detected signal is deterministically used
for phase evaluation, without post-selection. Super-resolution of
these two signals was demonstrated, up to 144 times better than
the Rayleigh limit. In addition, these single-shot measurements
follow the SNL up to pulses of 200 photons on average, and
eventually reach a sensitivity 230 times better than the
wavelength. The parity resolution is better than that of $P_0$,
but the $P_0$ sensitivity is slightly better and prevails for
larger photon numbers. The main limiting factor for these
measurements is the visibility of the interferometer.


\begin{thebibliography}{1}

\bibitem{Meyer88}G. Meyer and N. M. Amer, Appl. Phys. Lett. \textbf{53}, 1045 (1988).



\bibitem{Giovannetti11}V. Giovannetti, S. Lloyd, and L. Maccone, Nature Photonics \textbf{5}, 222 (2011).

\bibitem{Aasi13}J. Aasi \textit{et al.}, Nature Photon. \textbf{7}, 613 (2013).

\bibitem{Dowling08}J. P. Dowling, Contemp. Phys. \textbf{49}, 125 (2008).


\bibitem{Kuzmich98}A. Kuzmich and L. Mandel, Quant. Semiclass. Opt. \textbf{10}, 493 (1998).

\bibitem{Mitchell04}M. W. Mitchell, J. S. Lundeen, and A. M. Steinberg, Nature \textbf{429}, 161 (2004).

\bibitem{Nagata07}T. Nagata, R. Okamoto, J. L. O'Brien, K. Sasaki, and S. Takeuchi, Science \textbf{316}, 726 (2007).

\bibitem{Afek10}I. Afek, O. Ambar, and Y. Silberberg, Science \textbf{328}, 879 (2010).



\bibitem{Gao10a}W. -B. Gao, C. -Y. Lu, X. -C. Yao, P. Xu, O. G\"{u}hne, A. Goebel, Y. -A. Chen, C. -Z. Peng, Z. -B. Chen, and J. -W. Pan, Nature Phys. \textbf{6}, 331 (2010).

\bibitem{Kothe}C. Kothe, G. Bj\"{o}rk, and M. Bourennane, Phys. Rev. A \textbf{81}, 063836 (2010).

\bibitem{Higgins}B. L. Higgins, D. W. Berry, S. D. Bartlett, H. M. Wiseman, and G. J. Pryde, Nature \textbf{450}, 393 (2007).

\bibitem{Migdall}C. F. Wildfeuer, A. J. Pearlman, J. Chen, J. Fan, A. Migdall and J. P. Dowling, Phys. Rev. A \textbf{80}, 043822 (2009).

\bibitem{Khoury}G. Khoury, H. S. Eisenberg, E. J. S. Fonseca, and D. Bouwmeester, Phys. Rev. Lett. \textbf{96}, 203601 (2006).

\bibitem{Resch07}K. J. Resch, K. L. Pregnell, R. Prevedel, A. Gilchrist, G. J. Pryde, J. L. O'Brien, and A. G. White, Phys. Rev. Lett. \textbf{98}, 223601 (2007).

\bibitem{Distante13}E. Distante, M. Je\v{z}ek, U. L. Andersen, Phys. Rev. Lett. \textbf{111}, 033603 (2013).

\bibitem{Gao10b}Y. Gao, P. M. Anisimov, C. F. Wildfeuer, J. Luine, H. Lee, and J. P. Dowling, J. Opt. Soc. Am. B \textbf{27}, A170 (2010).

\bibitem{Bollinger}J. J. Bollinger, W. M. Itano, D. J. Wineland, and D. J. Heinzen, Phys. Rev. A \textbf{54}, R4649 (1996).

\bibitem{Gerry00}C. C. Gerry, Phys. Rev. A \textbf{61}, 043811 (2000).

\bibitem{Gerry01}C. C. Gerry and R. A. Campos, Phys. Rev. A \textbf{64}, 063814 (2001).

\bibitem{Gerry10}C. C. Gerry and J. Mimih, Phys. Rev. A \textbf{82}, 013831 (2010).

\bibitem{Chiruvelli}A. Chiruvelli and H. Lee, J. Mod. Opt. \textbf{58}, 945 (2011).

\bibitem{Anisimov}P. M. Anisimov, G. M. Raterman, A. Chiruvelli, W. N. Plick, S. D. Huver, H. Lee, and J. P. Dowling, Phys. Rev. Lett. \textbf{104}, 103602 (2010).

\bibitem{Plick}W. N. Plick, P. M. Anisimov, J. P. Dowling, H. Lee, and G. S. Agarwal, New J. Phys. \textbf{12}, 11 (2010).

\bibitem{Dovrat}L. Dovrat, M. Bakstein, D. Istrati, A. Shaham, and H. S. Eisenberg, Opt. Express \textbf{20}, 2266 (2010).

\bibitem{DovratSim}L. Dovrat, M. Bakstein, D. Istrati, and H. S. Eisenberg, Phys. Scr. \textbf{T147}, 014010 (2012).

\bibitem{Waks03}E. Waks, K. Inoue, W. D. Oliver, E. Diamanti, and Y. Yamamoto, IEEE J. Sel. Top. Quant. Elect. \textbf{9}, 1502 (2003).

\bibitem{Lita08}A. E. Lita, A. J. Miller, and S. W. Nam, Opt. Express \textbf{16}, 3032 (2008).



\end{thebibliography}
\end{document}